\documentclass[aps,prb,preprint,showpacs]{revtex4-1}
\usepackage{graphicx,color}
\usepackage{dcolumn}
\usepackage{bm}
\usepackage{amsmath}
\usepackage{ulem}
\usepackage{epstopdf}
\begin{document}

\title{ 
Transport properties of iron at the Earth's core conditions: the effect of spin disorder 
}

\date{\today}

\author{V. Drchal, J. Kudrnovsk\'y}
\affiliation{Institute of Physics, Academy of Sciences of the
Czech Republic, CZ-182 21 Praha 8, Czech Republic}

\author{D. Wagenknecht, I. Turek }
\affiliation{Charles University, Faculty of Mathematics and Physics,
Department of Condensed Matter Physics, Ke Karlovu 5,
CZ-121 16 Praha 2, Czech Republic}

\author{S. Khmelevskyi}
\affiliation{Center for Computational Materials Science,
Institute for Applied Physics, Vienna University of Technology,
Wiedner Hauptstrasse 8, A-1040 Vienna, Austria}
\date{\today}

\begin{abstract}
The electronic and thermal transport properties of the Earth's 
core are crucial for many geophysical models such as the geodynamo 
model of the Earth's magnetic field and of its reversals. 
Here we show, by considering bcc-iron and iron-rich iron-silicon alloy 
as a representative of the Earth's core composition and applying the 
first-principles modeling that the spin disorder at the Earth's core 
conditions not considered previously 
provides an essential contribution, of order 20~$\mu\Omega$\,cm, 
to the electrical resistivity. 
This value is comparable in magnitude with the electron-phonon and with 
the recently estimated electron-electron scattering contributions. 
The origin of the spin-disorder resistivity (SDR) consists in the existence 
of fluctuating local moments that are stabilized at high temperatures by 
the magnetic entropy even at pressures at which the ground state of iron 
is non-magnetic. 
We find that electron-phonon and SDR contributions are not additive 
at high temperatures.
We thus observe a large violation of the Matthiessen rule, not common in
conventional metallic alloys at ambient conditions.
\end{abstract}


\maketitle

\section{Introduction}

The temperature dependence of the resistivity is one of the most 
important properties of metals.
At ambient conditions, the resistivity of metals and their alloys
consists of three contributions: 
(i) the residual resistivity $\rho_{\rm imp}$ which is due to 
the scattering of conduction electrons on impurities and other 
structural defects with a very weak temperature dependence; 
(ii) the phonon contribution $\rho_{\rm ph}$, and 
(iii) the contribution $\rho_{\rm mag}$ which is due to the scattering 
on magnetic fluctuations in ferromagnetic metals and in alloys with 
local moments.
In an ideal ferromagnet $\rho_{\rm imp}=0$ while $\rho_{\rm ph}$ 
varies linearly with temperature $T$  above the Debye temperature and usually 
even below it down to fairly low temperatures \cite{rho-ph}.
The contribution due to magnetic fluctuations reaches its maximum at
the Curie temperature, $T_{\rm c}$, and then remains constant.
Its value corresponding to scattering of charge carriers on static disordered
moments is called the spin-disorder resistivity (SDR), $\rho_{\rm SDR}$, 
\cite{sdr-kl,sdr-our}.
The SDR is an important feature of ferromagnetic metals and the 
bcc-iron at ambient conditions is a textbook example with 
$\rho_{\rm mag}$ at $T_{\rm c}$ of about 80~$\mu\Omega$\,cm, four times larger
than its phonon part \cite{rho-exp}.
It should be noted that the temperature dependence of $\rho_{\rm mag}$ 
below $T_{\rm c}$ obeys reasonably well the $T^2$ law.

At the Earth's core conditions, i.e., at pressures about 350~GPa and 
temperatures 5000~K $-$ 6000~K, one naturally expects dominating scattering
on phonons, $\rho_{\rm ph}$, on which many authors studying resistivity 
have concentrated. 
We refer readers to a recent extensive review \cite{rho-rev}.
We mention in particular studies of Pozzo and Alf\`e \cite{rho-md} which
employ {\it ab initio} molecular dynamics simulations \cite{rho-mdth}.
Their estimate of $\rho_{\rm ph}$ about 50~$\mu\Omega$\,cm 
seems to agree reasonably well with the result of a very recent 
measurement employing the laser-heated diamond-anvil cell
\cite{rho-ph-dac}.
The estimated value of $\rho_{\rm ph}$ is few times smaller than the 
traditional estimates based on the extrapolation of the shock compression 
data \cite{rho-shock}.
Possible effects of electron correlations under Earth's core
conditions were investigated recently  \cite{pour13,vekil,rho-dmft}.
There are indications of a Fermi-liquid behavior of hcp and fcc phases
and a non-Fermi-liquid behavior of the bcc phase.
The resistivity, $\rho_{\rm ec}$, caused by electron correlations 
was calculated for hcp phase and the value $\rho_{\rm ec} = 16~\mu\Omega$\,cm 
was reported \cite{rho-dmft}.
It should be noted that a reliable experimental estimate of the 
resistivity at the Earth's core conditions is a highly demanding task.

In contrast to existing theoretical studies of electrical
resistivity due to phonons and electron correlations,
the role of spin polarization and magnetic fluctuations in
transport properties of iron at the Earth's core conditions
remains unexplored.
First-principles total energy calculations \cite{mag-earth} clearly
show that the long-range magnetic order does not exist.
On the other hand, authors of Ref.~\onlinecite{mag-earth} demonstrated
the existence of fluctuating Fe-local moment larger than 1~$\mu_{\rm B}$
for bcc Fe as well as for fcc Fe and hcp Fe using the classical model 
of local spin fluctuations (LSF).
One can understand this result as a stabilizing effect of the
magnetic entropy.
A similar effect exists also at ambient conditions, e.g., for fcc Ni 
in paramagnetic region \cite{mag-ni}.
In particular, the issue here was an estimate of the size of fluctuating
local Ni-moment just above $T_{\rm c}$, i.e., in the paramagnetic state.
A conventional description of the paramagnetic state using the disordered
local moment (DLM) approach \cite{dlm} gives a zero local moment.
The combination of the DLM with the fixed spin-moment (FSM) approach
\cite{mag-ni} yields the total energy that increases with the value 
of the fixed local moment $m_{\rm Ni}$ on Ni atoms, but the presence 
of magnetic entropy leads to a minimum of the free energy for the value of
$m_{\rm Ni}$ about 0.4~$\mu_{\rm B}$ in a very good agreement both with 
the LSF calculations \cite{ni-lsf} and the neutron diffuse scatterings
experiments.

The purpose of the present paper is a first-principles study of
the electrical resistivity of iron-based systems due to the spin
disorder relevant under the Earth's core conditions.
We focus not only on the SDR itself, but also address the combined
effect of several possible scattering mechanisms, namely, phonons,
spin disorder, and impurities.

\section{Theory}

The electronic structure was determined from first principles
by using the tight-binding linear muffin-tin orbital (TB-LMTO) 
method and the local spin-density approximation.
The effect of disorder was included in the framework of the coherent
potential approximation (CPA).
The CPA is used to include the effect of substitutional impurities 
(Si in the present work) and also to describe the paramagnetic state 
in terms of disordered local moments (DLM state)\cite{dlm}.
The effect of phonon disorder is treated within the multicomponent CPA. 
Statistical mechanics of disordered moments is based on the magnetic 
entropy proposed by Heine and Joynt\cite{magentr} and Grimvall\cite{Grimvall}.
We refer the reader to the Supplemental material \cite{SuMat} for details.

\subsection{Electron transport}   

The DLM-FSM method yields the selfconsistent potentials needed to 
calculate the SDR using the Kubo-Greenwood approach.
It should be noted that the DLM approach is closely related to the
conventional alloy theory employing the CPA \cite{dlm}.
Calculated SDR of fcc Ni at ambient conditions agrees well with 
the experiment (SDR is about 15~$\mu\Omega$\,cm) \cite{sdr-our}.
We use the same computational approach also for estimate of the 
SDR at the Earth's core conditions.

We need to determine $\rho_{\rm ph}$ for comparison with the SDR and 
to investigate if individual contributions to the total resistivity 
are additive -- in other words if the Matthiessen rule is valid.
We employ a simple yet quantitative model to include the effect of phonons.  
Their effect is accounted for by frozen random displacements of atomic 
positions.

A good agreement of calculated and measured $T$-dependent resistivity 
was obtained for $\rho_{\rm ph}$ at ambient conditions \cite{DW}
if the root-mean-square (r.m.s.) displacements $\sqrt{\langle u^{2} \rangle}$ 
for a given temperature are estimated from the Debye theory.
We refer the reader to the Supplemental material \cite{SuMat} for details.
It is important to note that results are also in a good agreement with
a more sophisticated {\it ab initio} molecular dynamics approach for
bcc Fe at ambient conditions \cite{rho-mdamb}.
However, a straightforward extension of the Debye theory to the Earth's
core condition is not justified as demonstrated recently \cite{rho-ph-dac}.
This theory yields large displacements, for example, for bcc Fe 
$\sqrt{\langle u^{2} \rangle} = 0.80$ bohr at 5500 K.
The relation between the temperature and displacements under
the Earth's core conditions can be estimated using other approaches:
(i) Lindemann's melting condition 
$\sqrt{\langle u^{2} \rangle} = \varrho \, r_{\rm m}$, where $\varrho$ is 
a constant and $r_{\rm m}$ is an interatomic distance \cite{gilvar};
 for bcc Fe it gives $\sqrt{\langle u^{2} \rangle} = 0.30$ bohr.
(ii) Molecular dynamics simulations \cite{msd-hcpFe}; the value 
$\sqrt{\langle u^{2} \rangle} = 0.59$ bohr was reported for hcp Fe.
The values estimated by these methods are rather scattered 
so we show calculated resistivities simply as a function of 
r.m.s. displacements.

The stable phase in the Earth's solid core is still under discussion.
The bcc-phase \cite{bcc-ec,bcc-ec2,ngeo} or the bcc-phase stabilized by 
sulphur or silicon impurities \cite{bcc-imp-ec}, treated in the present
study, are probable candidates, although hcp Fe is also possible, 
whereas fcc Fe seems less probable \cite{soderlind}.
The presence of impurities is needed to explain the density of the 
Earth's core which is smaller than if it consisted from pure iron.
The main aim of the present study is to understand the effect of spin 
fluctuations on transport so that a specific choice of the particular
phase is less important.

\subsection{Computational details}

All calculations were done using the scalar-relativistic 
TB-LMTO method while the effects of 
spin disorder and of atomic disorder were treated within the CPA \cite{book}.
The transport properties are determined using the Kubo-Greenwood
approach implemented in the framework of the TB-LMTO-CPA method \cite{kglmto}.
The $spdf$-basis set was employed.
We assumed the volume reduction 0.6:1 with respect to the ambient case 
(Wigner-Seitz radius is 2.250 bohr, or bcc lattice constant 2.418~\AA).
The change of the LMTO structure-constant matrix due to the atomic 
displacements can be recast into a change of the potential functions
of atoms located formally at the sites of the undistorted crystalline
lattice.
This model is treated in the multicomponent CPA \cite{rho-phth1,DW}. 
The reader can find computational details in the Supplemental material \cite{SuMat}.

\section{Results and discussion}

\subsection{Local moments}

\begin{figure}
\center \includegraphics[width=12cm]{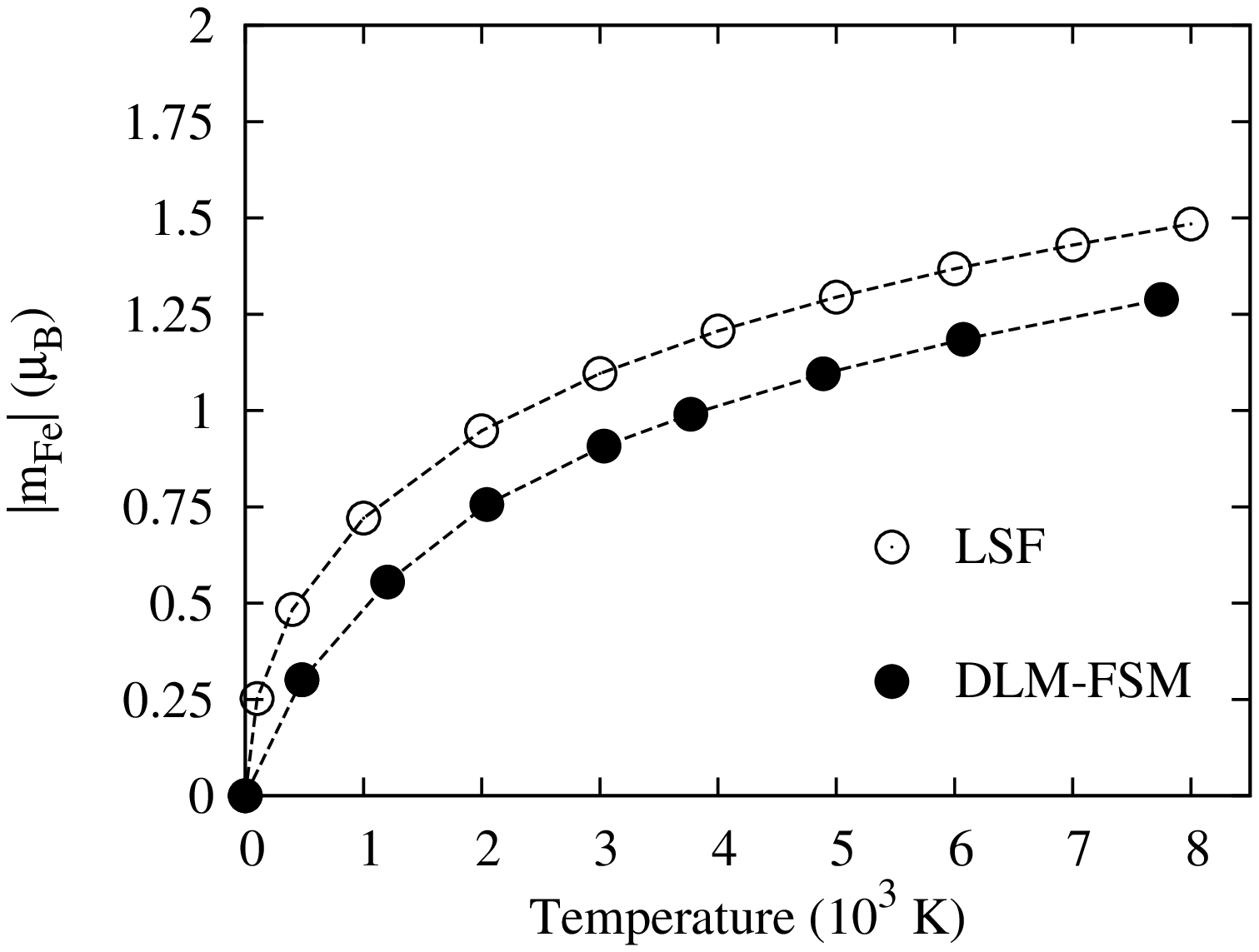}
\caption { Fluctuating values of the local Fe-moment $|m_{\rm Fe}|$ in bcc 
iron as a function of the temperature for the Earth's core atomic volume
(Wigner-Seitz radius is 2.25 bohr): (i) the LSF model \cite{mag-earth} 
(empty circles); and (ii) the present DLM-FSM model (full circles).
}
\label{fig1}
\end{figure}

The local Fe-moments as a function of temperature $T$ calculated using 
the DLM-FSM and the Heine-Joynt magnetic entropy \cite{magentr,Grimvall} are shown 
in Fig.~\ref{fig1} together with the values taken from Ref.
\onlinecite{mag-earth}.
We see a good agreement between the results of the LSF and DLM-FSM methods 
although the DLM-FSM moments are slightly smaller than the LSF ones.
We refer the reader to a recent paper, Ref.~\onlinecite{sk}, for 
a general discussion of both approaches.
Results indicate the existence of robust local Fe-moments 
stabilized by the magnetic entropy which monotonically increase
with temperature and which show a saturation at high temperatures. 
The values of order 1.1$-$1.3~$\mu_{\rm B}$ for temperatures 
5000$-$6000~K, typical for the Earth's core, are found.

\subsection{Spin-disorder resistivity}

\begin{figure}
\center \includegraphics[width=12cm]{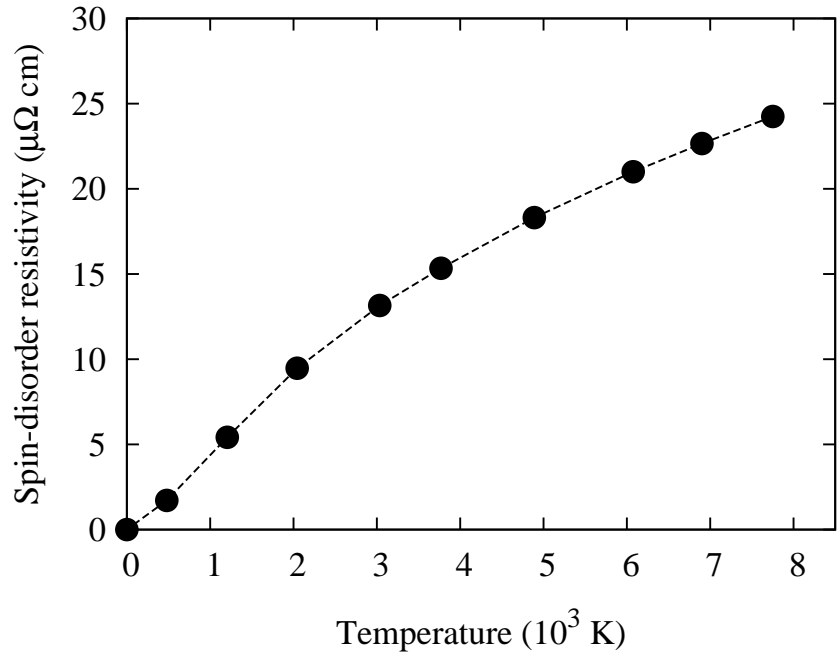}
\caption {Calculated SDR in bcc Fe as a function of the temperature
for the same atomic volume as in Fig.~\ref{fig1} with fluctuating
local moments obtained from the DLM-FSM model.
}
\label{fig2}
\end{figure}

The SDR's based on the DLM-FSM approach are shown in Fig.~\ref{fig2} as a function
of the temperature for the same volume as in Fig.~\ref{fig1}.
We see a monotonic increase of SDR with temperature while without the
stabilizing effect of the magnetic entropy the SDR is zero.
The SDR values of order 20~$\mu\Omega$\,cm are obtained for temperatures
in the Earth's core.
Such a value of the SDR contribution is about 2.5 times smaller 
than the phonon contribution, but larger than that coming
from the electron correlations \cite{rho-dmft}.
It is interesting to note that for small values of the r.m.s. deviations 
the calculated behavior is similar to that at ambient conditions.
In agreement with calculations using first-principles molecular
dynamics \cite{rho-rev} we observe saturation at high temperatures
corresponding to the Earth's core conditions.
It is interesting to compare the present result to that for
bcc Fe at ambient conditions.
First, the dominating contribution at $T_{\rm c}$=1050~K is the SDR,
about four-times larger than that due to phonons. 
Second, the Matthiessen rule in bcc Fe is obeyed quite well at 
ambient conditions, although even here the theory predicts some 
violation \cite{sdr-kirill}.

\subsection{Combined effect of spin disorder and phonons}

\begin{figure}
\center \includegraphics[width=12cm]{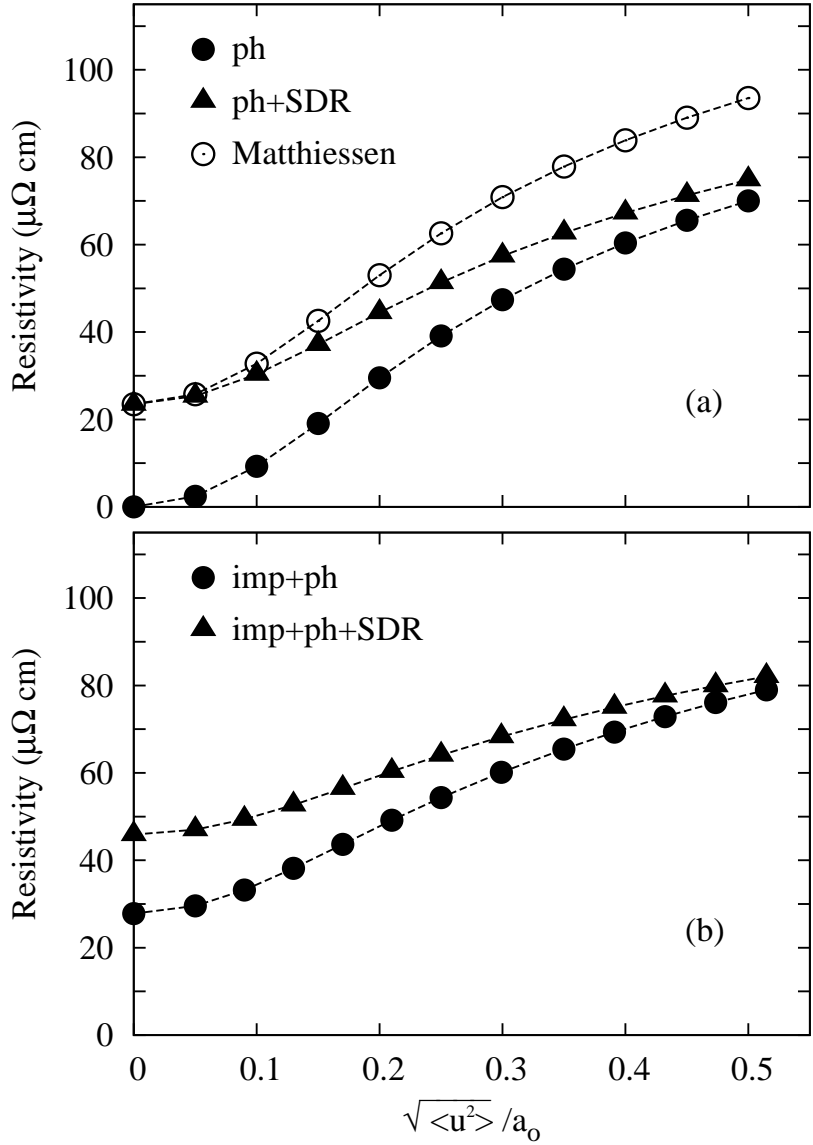}
\caption { The resistivity due to phonons as a function of the
r.m.s. displacement $\sqrt{\langle u^{2} \rangle}$
calculated for a model based on the
multicomponent CPA \cite{DW,rho-phth1} for the Earth's core
conditions: (a) Ideal bcc Fe with phonons only (filled
circles), with phonons and the SDR contribution for $T=5500$~K
calculated including both effects together on equal footing 
(filled triangles), and assuming the validity of the Matthiessen 
rule (empty circles); and (b) Disordered bcc Fe$_{0.92}$Si$_{0.08}$
alloy with phonons only (filled circles), with phonons and the SDR
contribution for $T=5500$~K including both effects together with
alloy disorder on equal footing (filled triangles). 
}
\label{fig3}
\end{figure}

Such result raises a natural question about the value of the total 
resistivity calculated assuming the validity of the Matthiessen rule, 
i.e., the additivity of $\rho_{\rm ph}$ and $\rho_{\rm SDR}$ or 
calculated when the effects of both phonons and spin disorder are 
included together on the same footing.
The temperature dependence of resistivity of systems with local moments
is very challenging problem even at ambient conditions
\cite{sdr-kirill,rho-phth1,DW}.
The situation is somewhat simpler if we can limit ourselves to the
paramagnetic region, i.e., to the case of the largest possible spin disorder 
at a given temperature, or better, for a specific local moment. 
We have chosen the moment size corresponding roughly to 5500~K, and
calculated total resistivities corresponding to various
r.m.s. displacements using the multicomponent CPA approach
\cite{rho-phth1,DW}.
We refer the reader to Supplemental material \cite{SuMat} for details.
The results are shown in Fig.~\ref{fig3}a together with the 
case with phonons only.
 For comparison, the total resistivity assuming the validity of
the Matthiessen rule, $\rho_{\rm M}  = \rho_{\rm ph} + \rho_{\rm SDR}$, 
is also shown. 
We employ $\rho_{\rm SDR}$ calculated at $T=5500$~K \cite{msd-hcpFe}.
The estimated value for $\sqrt{\langle u^{2} \rangle}$ is about
0.59 bohr, to which corresponds $\rho_{\rm ph}$ about 75~$\mu\Omega$\,cm.
This is about 50\% larger value than that calculated for hcp Fe in
Ref.~\onlinecite{rho-rev} or estimated experimentally although for
a smaller pressure (see Fig.~3 in Ref.~\onlinecite{rho-ph-dac}).
We ascribe this discrepancy to a simplified phonon model used here
(see Supplemental material \cite{SuMat} for details) and the missing 
knowledge of actual r.m.s. diplacement.
The most remarkable result is, however, a strong violation of the 
Matthiessen rule seen in Fig.~\ref{fig3}a: while the SDR is about 
20~$\mu\Omega$\,cm, the net increase of the $\rho_{\rm tot}$ is only 
2~$\mu\Omega$\,cm.
Even if we choose $\rho_{\rm ph}$ around 50~$\mu\Omega$\,cm, same
as that calculated in Ref.~\onlinecite{rho-rev}, which corresponds 
to the present r.m.s. displacement of 0.35 bohr, the net increase is
still only 9~$\mu\Omega$\,cm.

\subsection{Combined effect of spin disorder and other scattering mechanisms}

Now we discuss the combined effect of substitutional impurities, phonons, 
and spin fluctuations.
There are indications that nickel, silicon, oxygen, sulfur and other impurities exist
in the Earth's core \cite{bcc-imp-ec,rho-fesi,rho-fesio}.
Nickel is perhaps the most prominent one and the study of its effects was
so far limited to the non-magnetic case, see, e.g., Ref.~\onlinecite{cote}.
It is possible to include two or more magnetic elements (Fe and Ni),
but it would require significantly more demanding computations.
As we make no attempt to study the effect of impurities systematically
we selected a non-magnetic Si to illustrate the role of chemical disorder.
We take a disordered bcc Fe-rich alloy with silicon impurities, 
specifically Fe$_{0.92}$Si$_{0.08}$ as an example.
Alloys at Earth's core conditions were not studied within the LSF aproach.
We therefore extend the present DLM-FSM theory to substitutional alloys 
assuming zero moments on Si impurities.

The combined effect of phonons and Si impurities was studied using the 
first-principles molecular dynamics \cite{rho-fesi}.
The calculated resistivities were 51~$\mu\Omega$\,cm for hcp Fe and 
63~$\mu\Omega$\,cm for solid solution Fe$_{0.92}$Si$_{0.08}$ at Earth's 
core conditions.
We have evaluated resistivity due to the combined effect of
impurities, spin fluctuations, and phonons.
We assumed a random distribution of Si atoms on the bcc lattice 
(no clustering or local environment effects) and described the effect 
of phonons in terms of effective r.m.s. displacements similarly as in 
the case of pure iron.
The displacements of Fe and Si atoms in the alloy differ, but we
made no attempt to take this fact into account,
we just assumed an effective r.m.s. displacement common for both 
atomic species and determined resistivity as a function of it for the SDR 
calculated at $T=5500$~K.
The fluctuating moment in disordered alloy calculated using the 
DLM-FSM method is 1.114~$\mu_{\rm B}$.
It is slightly smaller than that calculated for pure bcc Fe at the 
same temperature, namely, 1.143~$\mu_{\rm B}$ which in turn is smaller
than that obtained by the LSF approach (see Fig.~\ref{fig1}). 

Results of our calculations are summarized in Fig.~\ref{fig3}b
in which the total resistivity including impurities, effect of phonons,
and spin disorder is shown together with a resistivity of the 
non-magnetic alloy Fe$_{0.92}$Si$_{0.08}$ due to impurities and phonons only.
In the latter case the value $\sqrt{\langle u^{2}\rangle}=0$
corresponds to impurity disorder alone (about 28~$\mu\Omega$\,cm).
The general conclusion is the same as for pure iron case: the
relative weight of the spin disorder in the total resistivity
decreases with temperature, or $\sqrt{\langle u^{2} \rangle}$ and
it approaches that with impurities and phonons only.
The calculated resistivities for pure iron and its alloy due to 
phonons only are 54 and 65~$\mu\Omega$\,cm (for 
$\sqrt{\langle u^{2} \rangle}=0.35$ bohr), respectively, which agrees well 
with the values 51 and 63~$\mu\Omega$\,cm obtained by the molecular 
dynamics simulations \cite{rho-fesi}.
A strong violation of the Matthiessen rule at Earth's core conditions was
verified also by using a finite-relaxation time model.
We refer the reader to the Supplemental material \cite{SuMat} for details.

With increasing temperature or with addition of further scattering 
mechanisms we observe a saturation of resistivity.
The contributions of individual scattering mechanisms to the total 
resistivity are not additive and the Matthiessen rule is violated.
This phenomenon is well-known and it is closely related to the Ioffe-Regel 
rule which states that the mean free path of charge carriers cannot
exceed the interatomic distance \cite{gunnar}.

Thermal and transport processes inside the Earth's core
are central to the notion of the geodynamo which is powered by the 
release of latent heat.
These processes are not independent, but mutually related 
(Wiedemann–-Franz law).
Our results show that due to the spin disorder the electrical 
resistivity can have higher value than expected on the basis
of previous calculations while the thermal conductivity will
be smaller. 
Although the properties of the liquid core cannot be directly
derived from the properties of the inner core, similar changes 
of outer core parameters can be anticipated.
One can thus expect that the effects of the spin disorder will 
finally act in favor of a stronger advection in the outer core.

\section{Conclusions} 

In summary, we have estimated from first principles a contribution to 
the resistivity of bcc Fe and of Fe-rich bcc FeSi alloy which is due 
to the presence of fluctuating spin moments stabilized at the Earth's 
core conditions by the magnetic entropy which was not considered 
in previous studies. 
The existence of fluctuating moments larger than 1~$\mu_{\rm B}$,
predicted earlier by the LSF approach, was confirmed 
by the present approach using the DLM-FSM method and the Heine-Joynt 
entropy also for FeSi alloys.
We used the multicomponent CPA method including vertex corrections 
to treat on equal footing three scattering mechanisms, namely, 
the scattering on spin disorder, on the atomic substitutional disorder,
and on atoms displaced from their equilibrium positions.

The estimated value of the SDR is about 20~$\mu\Omega$\,cm for bcc Fe.
Very rough estimates of other contributions, namely, $\rho_{\rm ph}$
from phonons and $\rho_{\rm ec}$ from electron correlations are
$\rho_{\rm ph} \approx 50 \mu\Omega$\,cm and $\rho_{\rm ec} \approx 16 \mu\Omega$\,cm.
The contribution from alloy disorder $\rho_{\rm dis}$ depends on the type 
and concentration of impurities, for bcc Fe$_{0.92}$Si$_{0.08}$ alloy we 
found $\rho_{\rm dis} \approx 28 \mu\Omega$\,cm.
All these contributions should be considered simultaneously as they
are not additive and a pronounced saturation is present.
This fact demonstrates a strong violation of the Matthiessen 
rule at the Earth's core conditions.
Consequently the microscopic origins of the transport properties of iron 
at the Earth's core conditions are markedly different from those at ambient 
conditions, where the violation of the Matthiessen rule is much weaker 
and, in addition, the SDR part dominates in the paramagnetic state.

The implications for geophysical phenomena are twofold: 
(i) the appearance of magnetic moments changes the electronic structure,
which can lead to modifications of all physical properties, and,
(ii) in particular, it brings a new contribution to electrical resistivity
and thus it can influence thermal conductivity.
These quantities are important parameters in the theory of the 
geodynamo and play a significant role in the thermal history of the
Earth.\cite{rho-ph-dac}

\begin{acknowledgments}
V.D., J.K., and I.T. acknowledge the financial support from 
the Czech Science Foundation (Grant No. 15-13436S), D.W. support from
the Grant  Agency of the Charles University (Grant No. 280815), and
S.Kh. acknowledges the Austrian FWF (SFB ViCoM F4109-N38). 
The National Grid Infrastructure MetaCentrum (Project No. LM2015042) 
is also acknowledged. 
This work was supported by The Ministry of Education, Youth and 
Sports from the Large Infrastructures for Research, Experimental 
Development and Innovations project "IT4Innovations National 
Supercomputing Center -- LM2015070".
V.D. and J.K. thank J. Kamar\'ad for useful discussions concerning 
high-pressure experiments. 

\end{acknowledgments}


\newpage

\newpage

\newpage

\end{document}